\documentclass{aastex} 
\usepackage{emulateapj5}

\usepackage{latexsym}

\shorttitle{Probes for the Hot Gaseous Halo of the LMC}
\shortauthors{Jaxon et al.}

\begin{document}

\title{Spectroscopic Classification of 42 LMC OB Stars:
Selection of Probes for the Hot Gaseous Halo of the LMC}

\author{Elizabeth G.\ Jaxon, Mart\'{\i}n A.\ Guerrero}
\affil{Astronomy Department, University of Illinois, 
        1002 W. Green Street, Urbana, IL 61801;
        jaxon@astro.uiuc.edu, mar@astro.uiuc.edu}
\author{J.\ Chris Howk\altaffilmark{1}}
\affil{Department of Physics and Astronomy, Johns Hopkins 
        University, 3400 N. Charles Street, Baltimore, MD 21218; 
        howk@pha.jhu.edu}
\author{Nolan R.\ Walborn}
\affil{Space Telescope Science Institute\altaffilmark{2}, 
        3700 San Martin Drive, Baltimore, MD 21218; 
        walborn@stsci.edu}
\author{You-Hua Chu}
\affil{Astronomy Department, University of Illinois, 
        1002 W. Green Street, Urbana, IL 61801;
        chu@astro.uiuc.edu}
\author{Bart P.\ Wakker}
\affil{Astronomy Department, University of Wisconsin,
       475 N. Charter Street, Madison, WI 53706;
       wakker@astro.wisc.edu}
\altaffiltext{1}{Visiting astronomer, Cerro Tololo Inter-American
         Observatory, National Optical Astronomy Observatories, 
        operated by the Association of Universities for Research 
       in Astronomy, Inc., under a cooperative agreement with the 
        National Science Foundation.}
\altaffiltext{2}{Operated by the Association of Universities for 
   Research in Astronomy, Inc., under NASA contract NAS5-26555.}

\begin{abstract}

Interstellar C~{\sc iv} absorption line studies of the hot gaseous 
halo of the Large Magellanic Cloud (LMC) have been hindered by 
non-ideal selections of early-type probe stars in regions where 
C$^{+3}$ can be produced locally via photoionization, fast stellar 
winds, or supernovae.  To observe stars outside such regions, precise 
spectral classifications of OB stars in the field are needed.  
Therefore, we have obtained medium-dispersion spectra of 42
early-type stars in the LMC that are distributed outside superbubbles 
or supergiant shells.  The spectral classification of these stars
is presented in this paper.  Nineteen of these program stars
have spectral types between B1 and O7, and are thus suitable probes
for interstellar C~{\sc iv} absorption line studies of the hot gaseous
halo of the LMC. 

\end{abstract}


\section{Introduction}

The existence of a hot gaseous halo (10$^6$ K) around the Milky Way was 
first suggested by \citet{spitzer56} to explain the confinement of 
\ion{H}{1} clouds at large distances from the Galactic plane.  
This proposed hot gaseous halo was later detected around the Galaxy 
and has been well studied via absorption line measurements of 
interstellar \ion{C}{4} and other highly-ionized species
\citep[e.g.,][]{SSL97}.  
The hot gas around a galaxy is most likely injected into the halo 
as a result of stellar winds and supernova explosions.  
The study of its physical properties and distribution in galactic 
halos provides valuable information about the physical structure 
and evolution of the interstellar medium (ISM).  

The disk-halo connection and the distribution of hot halo gas over 
the disk are difficult to study in the Galaxy because of our 
disadvantageous position in the disk plane.
The Large Magellanic Cloud (LMC), on the other hand, provides an
excellent laboratory for the study of a hot gaseous halo, because of 
its small inclination angle \citep[30$^\circ$--40$^\circ$;][]{we97},
proximity \citep[distance = 50 kpc;][]{feast99}, and low 
extinction \citep[typical A$_{\rm V} \ll$ 0.5 mag;][]{CD86}.
However, the hot gaseous halo of the LMC has not been adequately studied.
Earlier observations were possible only for the most luminous OB probe
stars.  These stars and their neighboring stars can produce C$^{+3}$ 
locally via photoionization, stellar winds, and supernovae, as 
illustrated by \citet{W84} with the hot stars in the Carina Nebula.
Therefore, the observed C$^{+3}$ may not represent the halo \citep{chuetal94}.  

To avoid these problems, suitable probe stars for further studies of
the hot halo in the LMC must 
(1) have spectral types later than O7, so they cannot photoionize 
    C$^{+2}$, 
(2) have spectral types earlier than B1, so they are bright enough 
    in the ultraviolet to obtain high S/N data, and 
(3) be located outside of supergiant shells or away from early O stars 
    that may produce confusing \ion{C}{4} lines.  
The existing catalogs \citep[e.g.,][]{S69,R78} list many stars that 
may meet these criteria, but their spectral types have been broadly 
categorized as ``OB", given the difficulty of classifying 
late-O/early-B stars with photographic objective prism plates.  
Accurate spectral classification of selected stars outside OB 
associations has been provided by, e.g., \citet{W77}, \citet{C86}, 
\citet{F91}, \citet{M95}, and \citet{W95}, but only a small number 
of stars meet these criteria.
Furthermore, these stars do not provide an even coverage across the 
disk of the LMC for a thorough mapping of its hot gaseous halo.
Therefore, a detailed spectral classification of more stars in
strategically selected regions in the LMC is needed.

We have obtained medium-dispersion spectroscopic observations in the 
3860--4790 {\AA} wavelength range of a sample of 42 Sanduleak stars 
that are outside superbubbles and supergiant shells, and selected 
from widespread fields in the LMC.  The observations have allowed us
to narrow the spectral classification of these stars.  The results 
are reported in this paper.  We describe the observations and data 
reduction in \S2, discuss the classification method in \S3, and 
present the improved spectral classifications in \S4.

\section{Observations and Reduction}

The observations were obtained using the Ritchey-Chr\'etien 
Spectrograph on the 1.5m telescope at the Cerro Tololo Inter-American 
Observatory on 1998 January 5--8.  
The $\#35$ grating was used in the second order to provide a 
reciprocal dispersion of 50.7 {\AA}~mm$^{-1}$.  
A CuSO$_4$ filter was used to block the first spectral order.  
The spectra were recorded with a Loral 1K chip.
The $15\mu$ pixel size corresponded to 0.76 {\AA}~pix$^{-1}$.
A $2\arcsec$ slit width was used, and the resultant spectral 
resolution was 2.4 \AA\ FWHM.
The wavelength coverage of the spectra was 3860--4790 {\AA}.  
The typical $S/N$ is above 80 per resolution element over the 
spectral range.
The journal of observations is given in Table~1.  
There we list the star names from Sanduleak's catalog, B magnitudes 
from SIMBAD, and the total integration times.  
In addition to the 42 program stars, we have obtained spectra for
20 Galactic O and B spectral classification standard stars selected
from the compilation by \citet{G89}.  The names, spectral 
classifications, and references of these stars are listed in Table~2.  


The observations were reduced using standard IRAF\footnote{
IRAF is distributed by the National Optical Astronomy Observatories, 
which are operated by the Association of Universities for Research in 
Astronomy, Inc., under cooperative agreement with the National Science 
Foundation.} packages.  
All spectra were bias and dark corrected and multiple exposures were 
combined to remove cosmic-ray events.  
The spectra were flat-fielded using observations of a quartz lamp, and 
the wavelength scale was calibrated using observations of a He--Ar 
comparison lamp.

\section{Classification of OB Stars}

The spectra of all standard and program stars were normalized and 
plotted with the same scale for comparison.  The spectral classification
of the program stars is carried out by matching diagnostic spectral 
features with those of the standard stars.  
It should be borne in mind that the LMC program stars have a lower 
metallicity than the Galactic standard stars, and the strengths of their 
metal lines can differ for the same spectral type.
Thus, the spectral classification is mostly based on line ratios.
We have followed the spectral classification criteria outlined by
\citet{WF90} and \citet{F91}, as summarized below.

A main classification criterion in the OB range is the strength 
of the He~{\sc ii} $\lambda\lambda$4200,4541 absorption lines.  
The He~{\sc ii} $\lambda$4541 line is visible in spectra of O
stars and even B0 stars.
For O stars, the detailed spectral type is further defined by the 
relative intensity of the He~{\sc ii} $\lambda\lambda$4200,4541 to 
the He~{\sc i} $\lambda\lambda$4026,4144,4387,4471 lines.  
At spectral types O6--O8, the luminosity classes are diagnosed by
the presence of N~{\sc iii} $\lambda\lambda$4634,4640,4642 emission 
lines and whether the He~{\sc ii} $\lambda$4686 line is in emission 
or absorption.
At later spectral types, O8--O9.5, the Si~{\sc iv} $\lambda$4089 
to H$\delta$ ratio is used to determine luminosity classes,
with large ratios for supergiants.

For early B stars, the spectral type is determined by the relative 
strengths of the Si~{\sc iv} $\lambda$4089, Si~{\sc iii} $\lambda$4552, 
and Si~{\sc ii}  $\lambda\lambda$4128,30 lines.  
For late B5--B9 stars, the Mg~{\sc ii} $\lambda$4481 to He~{\sc i} 
$\lambda$4471 ratio becomes an effective diagnostic.  
The luminosity class for B type stars is mainly determined by the relative 
strengths of Si~{\sc iii} and Si~{\sc iv} compared to He~{\sc i}.

\section{Results}

The normalized spectra of the program stars in the wavelength range 
$3880-4775$ {\AA} are displayed in Figure~1 with a separation of
0.5 continuum flux units between spectra.
The spectra have been ordered according to their spectral 
classification, from earlier to later spectral types, and by 
luminosity class.
This order is chosen to facilitate easy viewing of the progressive
spectral variations from one spectral type to the next.

The spectral classification of the program stars is summarized in 
Table\,3.  The stars have been arranged according to the order in 
Sanduleak's catalog.  
Columns 1 and 2 give the Sanduleak and HDE numbers, 
columns 3 and 4 the J2000.0 coordinates as measured in the 
Digitized Sky Survey\footnote{The Digitized Sky Survey was produced 
at the Space Telescope Science Institute under US government grant 
NAGW-2166 based on photographic data obtained using the UK Schmidt 
Telescope at Siding Spring, Australia, and the Oschin Schmidt 
Telescope on Palomar Mountain.}, 
and columns 5 and 6 the $B$ and $V$ magnitudes as provided by the 
SIMBAD database.  
Finally, our spectral classification is given in column 7.
Two stars have uncertain luminosity classes, and they are
noted with a ``:'' following the classification.

After we have classified all program stars, we find that three 
of them actually have been accurately classified previously --
Sk$-65\arcdeg22$ (O6\,Iaf+, Walborn 1977),  and Sk$-66\arcdeg78$ 
(B1\,Ia) and Sk$-68\arcdeg171$ (B0.7\,Ia, Fitzpatrick 1991).
Our independent classifications are consistent with these previous
results within the uncertainty of human judgement.

Finally, Sk$-69\arcdeg59$ shows conflicting spectral features that 
lead to a composite classification.  
The presence of He~{\sc ii} suggests a spectral type B0 or earlier, 
while the low ratios of Si~{\sc iv} $\lambda$4089 to 
Si~{\sc iii} $\lambda$4552 and Si~{\sc iv} $\lambda$4089 to 
He~{\sc i} $\lambda\lambda$4121,4144  are consistent with 
a spectral type B1; therefore, Sk$-69\arcdeg59$ may have a composite 
spectrum of O and B1 stars.  

Of the 42 program stars, 19 have spectral types between B1 and O7.
As the program stars have been selected to sample different regions
of the LMC and to avoid environments that are expected to possess
local hot gas, these 19 stars are suitable probes for interstellar 
C~{\sc iv} absorption line studies of the hot gaseous halo of the LMC.
High-dispersion spectroscopic observations of these probe stars in the 
C~{\sc iv} lines using the Space Telescope Imaging Spectrograph onboard 
the {\it Hubble Space Telescope} and in the O~{\sc vi} lines using the 
{\it Far Ultraviolet Spectroscopic Explorer} would help us understand 
better the distribution and physical conditions of the hot gaseous halo 
of the LMC.

\acknowledgements
JCH acknowledges support from NASA Long Term Space Astrophysics grant
NAG5-3485 through the Johns Hopkins University and from a NASA
Graduate Student Researcher Fellowship under grant number NGT-5-50121
through the University of Wisconsin-Madison.
BW acknowledges NASA grants NAG 5-9024 and NAG 5-9179.

\clearpage


\begin{deluxetable}{llc|llc|llc}
\tablenum{1}
\tablewidth{40pc}
\tablecaption{Journal of Observations}
\tablehead{
\multicolumn{1}{c}{Sk} &
\multicolumn{1}{c}{$B$} & 
\multicolumn{1}{c}{Exp. Time} & 
\multicolumn{1}{c}{Sk} &
\multicolumn{1}{c}{$B$} & 
\multicolumn{1}{c}{Exp. Time} & 
\multicolumn{1}{c}{Sk} &
\multicolumn{1}{c}{$B$} & 
\multicolumn{1}{c}{Exp. Time} \\
\multicolumn{1}{l}{} &
\multicolumn{1}{c}{[mag]} & 
\multicolumn{1}{c}{[s]} & 
\multicolumn{1}{l}{} &
\multicolumn{1}{c}{[mag]} & 
\multicolumn{1}{c}{[s]} & 
\multicolumn{1}{l}{} &
\multicolumn{1}{c}{[mag]} & 
\multicolumn{1}{c}{[s]} 
}

\startdata
$-$65\arcdeg 1   & ~12.4   & 2100 & $-$67\arcdeg 249 & ~12.4   & 3000 & $-$69\arcdeg 65  & ~12.6   & 2400 \nl
$-$65\arcdeg 22  & ~11.9   & 1500 & $-$68\arcdeg 7  & ~12.8   & 3000 & $-$69\arcdeg 68  & ~12.5   & 2100 \nl
$-$65\arcdeg 41  & ~12.7   & 1800 & $-$68\arcdeg 23  & ~13.0   & 3000 & $-$69\arcdeg 76  & ~11.8   & 2700 \nl
$-$65\arcdeg 66  & ~12.9   & 3000 & $-$68\arcdeg 46  & ~12.4   & 2700 & $-$69\arcdeg 97  & ~12.8   & 2400 \nl
$-$66\arcdeg 6   & ~12.6   & 2400 & $-$68\arcdeg 75  & ~12.0   & 2100 & $-$69\arcdeg 120 & ~12.6   & 3000 \nl
$-$66\arcdeg 78  & ~12.2   & 2700 & $-$68\arcdeg 147 & ~12.9   & 3000 & $-$69\arcdeg 305 & ~13.0   & 3000 \nl 
$-$67\arcdeg 4   & ~12.8   & 3000 & $-$68\arcdeg 166 & ~12.6   & 2800 & $-$70\arcdeg 25  & ~12.9   & 3000 \nl
$-$67\arcdeg 45  & ~12.9   & 3000 & $-$68\arcdeg 171 & ~11.9   & 1560 & $-$70\arcdeg 84  & ~12.7   & 3000 \nl
$-$67\arcdeg 46  & ~12.3   & 2100 & $-$68\arcdeg 175 & ~12.2   & 3000 & $-$70\arcdeg 85  & ~12.2   & 2700 \nl
$-$67\arcdeg 54  & ~~$\dots$ & 2100 & $-$68\arcdeg 180 & ~12.8   & 3000 & $-$70\arcdeg 96  & ~12.5   & 3000 \nl
$-$67\arcdeg 57  & ~12.6   & 1800 & $-$69\arcdeg 9   & ~12.4   & 3000 & $-$70\arcdeg 106 & ~12.9   & 3000 \nl
$-$67\arcdeg 76  & ~12.3   & 2400 & $-$69\arcdeg 45  & ~12.2   & 2400 & $-$71\arcdeg 7   & ~12.8   & 3200 \nl
$-$67\arcdeg 91  & ~12.3   & 2700 & $-$69\arcdeg 51  & ~12.5   & 4000 & $-$71\arcdeg 9   & ~12.5   & 3000 \nl 
$-$67\arcdeg 247 & ~12.2   & 2100 & $-$69\arcdeg 59  & ~12.0   & 2400 & $-$71\arcdeg 11  & ~12.6   & 3600 \nl
\enddata
\end{deluxetable}
 

\begin{deluxetable}{lll}
\tablenum{2}
\tablewidth{20pc}
\tablecaption{Spectral Classification of Standard Stars}
\tablehead{
\multicolumn{1}{c}{Star} & 
\multicolumn{1}{l}{~~Spectral Type~~} & 
\multicolumn{1}{l}{Reference~~} }

\startdata

~~29~CMa           & ~ O7\,Ia:fp var    & ~~~~1   \nl 
~~HD~34656         & ~ O7\,II (f)       & ~~~~2   \nl 
~~$\xi$~Per        & ~ O7\,III (n)((f)) & ~~~~3   \nl 
~~HD~112244        & ~ O8.5\,Iab (f)    & ~~~~3,4 \nl
~~$\lambda$~Ori    & ~ O8\,III ((f))    & ~~~~3,4 \nl 
~~$\tau$~CMa       & ~ O9\,II           & ~~~~3   \nl 
~~$\iota$~Ori      & ~ O9\,III          & ~~~~4   \nl
~~$\delta$~Ori     & ~ O9.5\,II         & ~~~~4   \nl 
~~$\zeta$~Ori      & ~ O9.7\,Ib         & ~~~~4   \nl 
~~$\epsilon$~Ori   & ~ B0\,Ia           & ~~~~4   \nl
~~HD~43818         & ~ B0\,II           & ~~~~5,6 \nl 
~~$\kappa$~Ori     & ~ B0.5\,Ia         & ~~~~3,4 \nl 
~~$\epsilon$~Per   & ~ B0.7\,III        & ~~~~3   \nl
~~$\epsilon$~CMa   & ~ B1.5\,II         & ~~~~4   \nl 
~~$\chi^2$~Ori     & ~ B2\,Ia           & ~~~~4   \nl 
~~$\gamma$~Ori     & ~ B2\,III          & ~~~~4   \nl
~~$o^2$~CMa        & ~ B3\,Ia           & ~~~~4   \nl 
~~$\iota$~CMa      & ~ B3\,Ib           & ~~~~4   \nl 
~~HD~21483         & ~ B3\,III          & ~~~~6,7 \nl  
~~$\eta$~CMa       & ~ B5\,Ia           & ~~~~4   \nl

\enddata
\tablerefs{(1) Walborn 1973; (2) Walborn 1972; (3) Walborn 1971; 
(4) Walborn \& Fitzpatrick (1990); (5) Morgan, Code, \& Whitford 
(1955); (6) Walborn (1976); (7) Rountree Lesh (1968)}
\end{deluxetable}

\begin{deluxetable}{lccclll}
\tablenum{3}
\tablewidth{36pc}
\tablecaption{Spectral Classification of Program Stars in the LMC}
\tablehead{
\multicolumn{1}{c}{Sk} & 
\multicolumn{1}{c}{HDE} & 
\multicolumn{1}{c}{R.\ A.} & 
\multicolumn{1}{c}{Dec.} & 
\multicolumn{1}{c}{$B$} & 
\multicolumn{1}{c}{$V$} & 
\multicolumn{1}{l}{Spectral Type} \\
\multicolumn{1}{l}{} &
\multicolumn{1}{l}{} &
\multicolumn{2}{c}{(J2000.0)} & 
\multicolumn{1}{c}{[mag]} & 
\multicolumn{1}{c}{[mag]} & 
\multicolumn{1}{c}{}
}

\startdata
$-$65\arcdeg 1   & $\dots$ & 04 54 06.8 & $-$65 35 27  & ~12.4  & ~12.5  & ~~B0.5 I     \nl 
$-$65\arcdeg 22  & 270952  & 05 01 23.3 & $-$65 52 37  & ~11.9  & ~12.1  & ~~O6   Iaf+  \nl 
$-$65\arcdeg 41  & $\dots$ & 05 19 05.7 & $-$65 40 06  & ~12.7  & ~12.9  & ~~B2   III   \nl
$-$65\arcdeg 66  & $\dots$ & 05 32 32.6 & $-$65 51 42  & ~12.9  & ~13.1  & ~~B0.5 II    \nl
                 &         &            &              &        &        &              \nl
$-$66\arcdeg 6   & $\dots$ & 04 53 35.1 & $-$66 56 21  & ~12.6  & ~12.7  & ~~B0.7 II    \nl 
$-$66\arcdeg 78  & $\dots$ & 05 23 30.4 & $-$66 42 14  & ~12.2  & ~12.2  & ~~B1.5 I     \nl 
                 &         &            &              &        &        &              \nl
$-$67\arcdeg 4   & $\dots$ & 04 49 52.0 & $-$67 41 17  & ~12.8  & ~13.0  & ~~O9   Ib    \nl 
$-$67\arcdeg 45  & $\dots$ & 05 06 18.8 & $-$67 41 49  & ~12.9  & ~13.0  & ~~B2   II    \nl
$-$67\arcdeg 46  & $\dots$ & 05 07 01.6 & $-$67 37 31  & ~12.3  & ~12.4  & ~~B1.5 I     \nl
$-$67\arcdeg 54  & $\dots$ & 05 10 52.0 & $-$67 10 29 &~~$\dots$& ~12.5  & ~~B1.5 I:    \nl
$-$67\arcdeg 57  & $\dots$ & 05 11 52.2 & $-$67 10 01  & ~12.6  & ~12.3  & ~~B2   III   \nl
$-$67\arcdeg 76  & $\dots$ & 05 20 05.8 & $-$67 21 12  & ~12.3  & ~12.4  & ~~B0   I     \nl 
$-$67\arcdeg 91  & $\dots$ & 05 23 14.7 & $-$67 26 07  & ~12.3  & ~12.3  & ~~B8-9 I     \nl 
$-$67\arcdeg 247 & 270037  & 05 42 09.2 & $-$67 17 29  & ~12.2  & ~12.2  & ~~B2.5 II    \nl
$-$67\arcdeg 249 & 270060  & 05 43 03.1 & $-$67 22 34  & ~12.4  & ~12.4  & ~~B1.5 I     \nl
                 &         &            &              &        &        &              \nl
$-$68\arcdeg 7   & $\dots$ & 04 53 27.0 & $-$68 30 00  & ~12.8  & ~12.9  & ~~B0.5 I     \nl
$-$68\arcdeg 23  & $\dots$ & 05 00 49.3 & $-$68 07 11  & ~13.0  & ~12.8  & ~~B3   I     \nl
$-$68\arcdeg 46  & $\dots$ & 05 06 14.8 & $-$68 33 03  & ~12.4  & ~12.4  & ~~O9.5 Ib    \nl 
$-$68\arcdeg 75  & 269463  & 05 23 28.4 & $-$68 12 25  & ~12.0  & ~12.0  & ~~B1   I     \nl
$-$68\arcdeg 147 & $\dots$ & 05 40 56.2 & $-$68 30 37  & ~12.9  & ~13.0  & ~~B2   II    \nl 
$-$68\arcdeg 166 & $\dots$ & 05 48 19.4 & $-$68 14 48  & ~12.6  & ~12.8  & ~~O8   II    \nl 
$-$68\arcdeg 171 & 270220  & 05 50 23.1 & $-$68 11 26  & ~11.9  & ~12.0  & ~~B0.7 I     \nl 
$-$68\arcdeg 175 & $\dots$ & 05 52 14.3 & $-$68 04 07  & ~12.2 &~~$\dots$& ~~B2   I/II  \nl
$-$68\arcdeg 180 & $\dots$ & 05 55 55.6 & $-$68 13 46  & ~12.8  & ~13.0  & ~~O6   II(f) \nl
                 &         &            &              &        &        &              \nl
$-$69\arcdeg 9   & $\dots$ & 04 49 51.6 & $-$69 12 04  & ~12.4  & ~12.6  & ~~O6.5 III   \nl
$-$69\arcdeg 45  & $\dots$ & 04 56 51.2 & $-$69 18 50  & ~12.2  & ~12.4  & ~~B0.7 I     \nl
$-$69\arcdeg 51  & $\dots$ & 04 57 50.7 & $-$69 41 09  & ~12.5  & ~12.5  & ~~B1.5 I:    \nl
$-$69\arcdeg 59  & 268960  & 05 03 12.8 & $-$69 02 37  & ~12.0  & ~12.1  & ~~Comp. O(f) + B1 I? \nl 
$-$69\arcdeg 65  & $\dots$ & 05 08 12.4 & $-$69 13 55  & ~12.6  & ~12.7  & ~~B2.5 I/II  \nl
$-$69\arcdeg 68  & $\dots$ & 05 09 58.7 & $-$69 07 03  & ~12.5  & ~12.5  & ~~B1   II    \nl
$-$69\arcdeg 76  & 269215  & 05 13 38.1 & $-$69 18 03  & ~11.8  & ~11.9  & ~~O6.5 II(f) \nl 
$-$69\arcdeg 97  & $\dots$ & 05 18 18.1 & $-$69 45 51  & ~12.8  & ~12.8  & ~~B2   I     \nl
$-$69\arcdeg 120 & $\dots$ & 05 24 21.4 & $-$69 42 48  & ~12.6  & ~12.7  & ~~B0   I     \nl
$-$69\arcdeg 305 & $\dots$ & 05 54 12.9 & $-$69 29 57  & ~13.0  & ~13.0  & ~~B5   Ia    \nl
                 &         &            &              &        &        &              \nl
$-$70\arcdeg 25  & $\dots$ & 04 58 19.8 & $-$70 13 05  & ~12.9  & ~13.1  & ~~B1.5 II    \nl 
$-$70\arcdeg 84  & $\dots$ & 05 16 13.7 & $-$70 34 30  & ~12.7  & ~12.8  & ~~B0.5 I     \nl 
$-$70\arcdeg 85  & 269314  & 05 17 05.7 & $-$70 19 24  & ~12.2  & ~12.3  & ~~B0   I     \nl
$-$70\arcdeg 96  & $\dots$ & 05 29 23.9 & $-$70 14 03  & ~12.5  & ~12.7  & ~~B0.7 III   \nl 
$-$70\arcdeg 106 & $\dots$ & 05 40 09.9 & $-$70 27 04  & ~12.9  & ~12.8  & ~~B3   I     \nl
                 &         &            &              &        &        &              \nl
$-$71\arcdeg 7   & $\dots$ & 05 06 57.2 & $-$71 13 24  & ~12.8  & ~12.9  & ~~B1   I     \nl 
$-$71\arcdeg 9   & $\dots$ & 05 07 35.6 & $-$71 12 28  & ~12.5  & ~12.6  & ~~B0   II    \nl
$-$71\arcdeg 11  & $\dots$ & 05 07 39.8 & $-$71 01 38  & ~12.6  & ~12.7  & ~~B0.5 I     
\enddata
\end{deluxetable}

\clearpage

\begin{figure}
\epsscale{0.9}
\centerline{\plotone{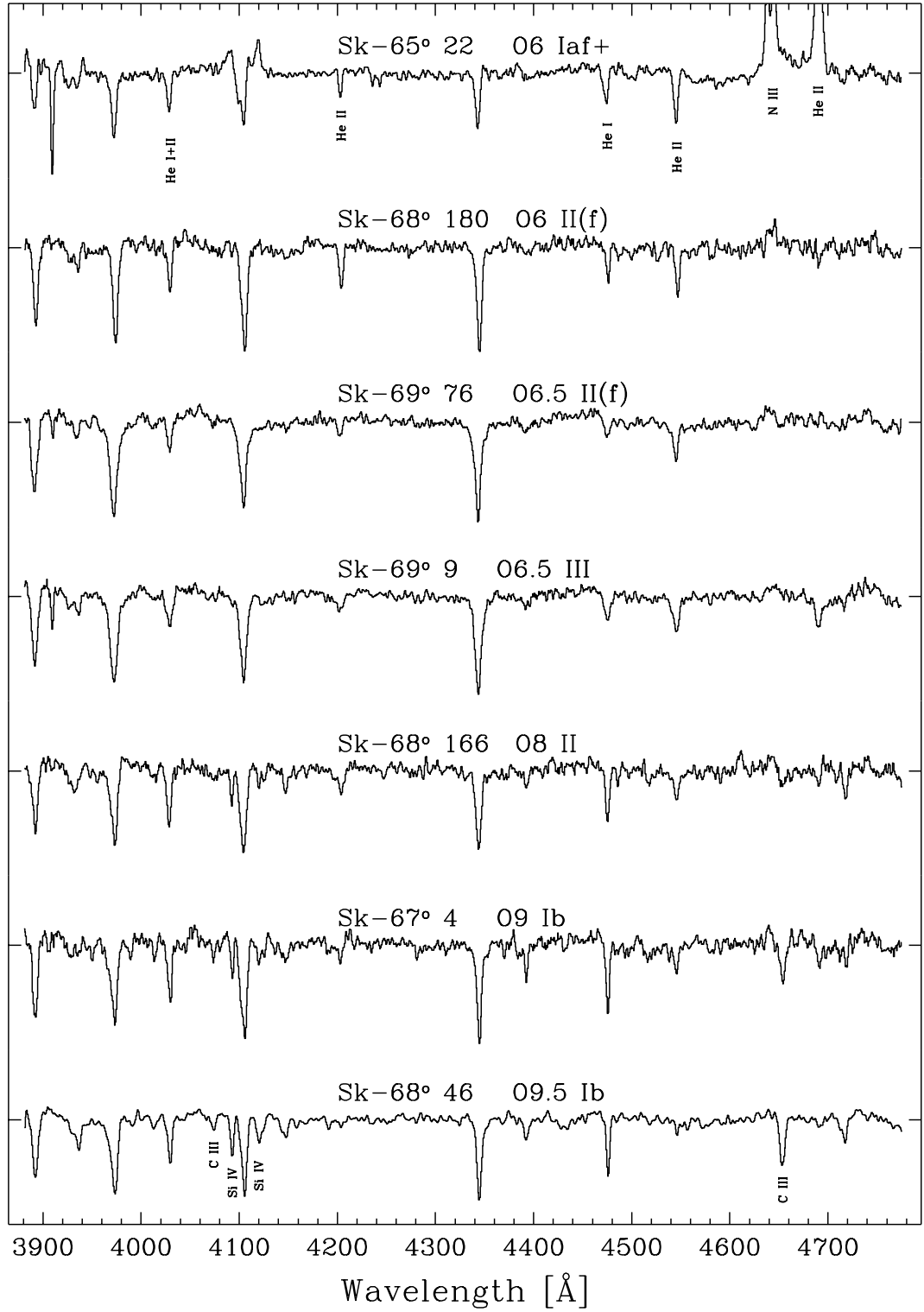}}
\caption{Normalized spectra of 42 early-type stars in the LMC.  The ordinate
ticks are separated by 0.5 continuum flux units.  The name and spectral
type are labeled above each spectrum.  The relevant diagnostic lines are
also marked.  See Table~1 for the exposure times of these spectra, and 
Table~3 for the coordinates and magnitudes of these stars.}
\end{figure}

\begin{figure}
\centerline{\plotone{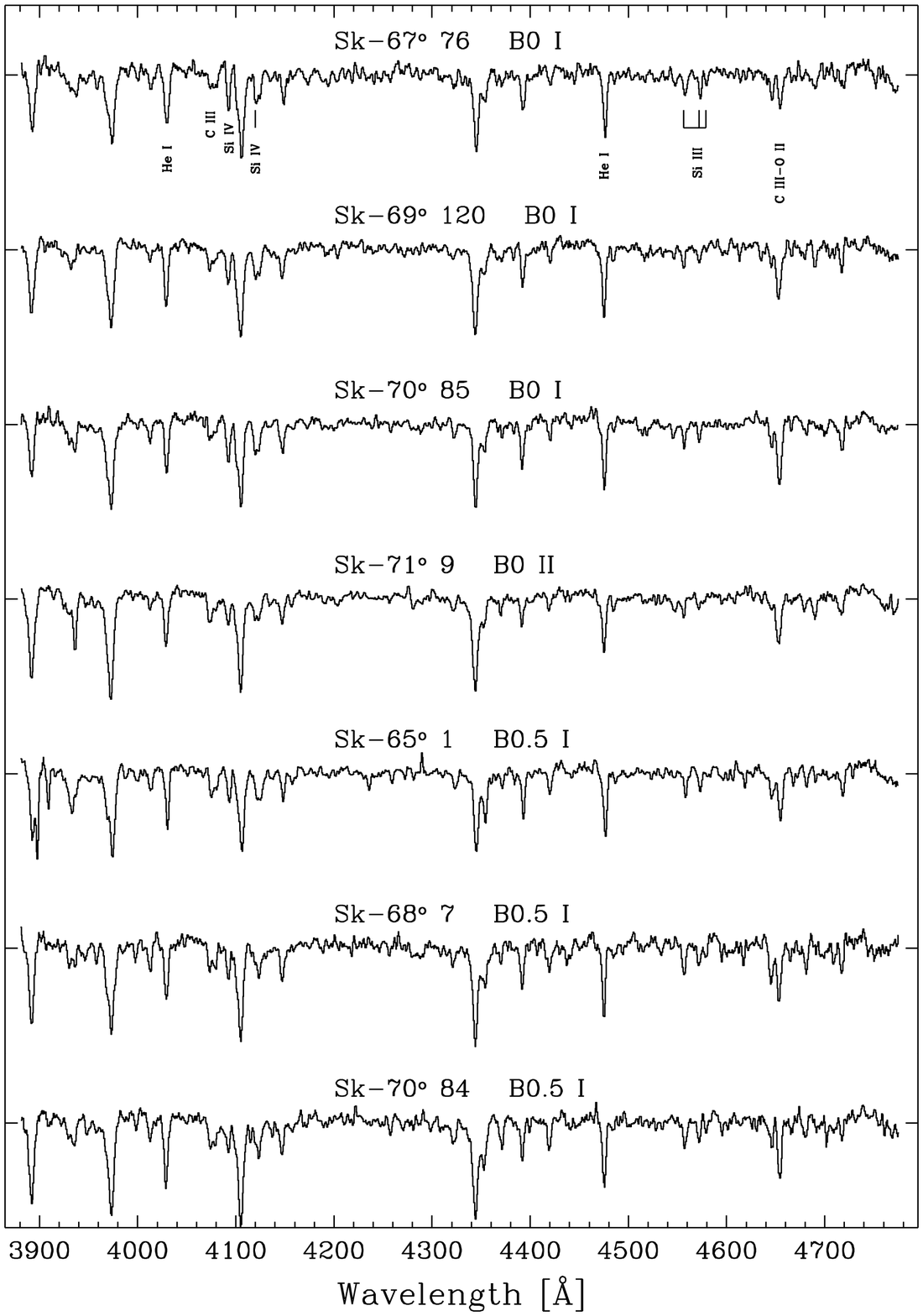}}
\end{figure}	     
		     
\begin{figure}	     
\centerline{\plotone{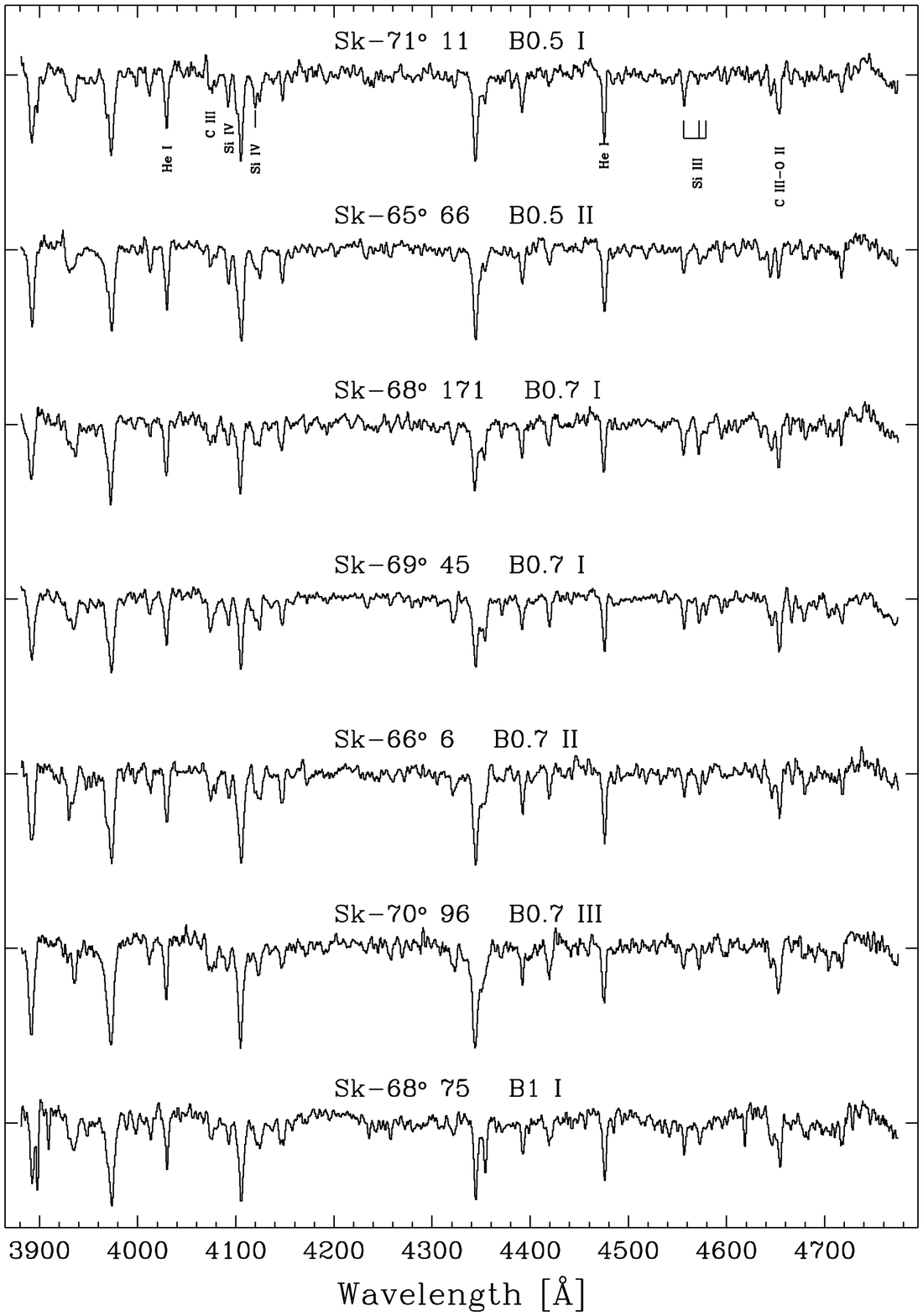}}
\end{figure}	     
		     
\begin{figure}	     
\centerline{\plotone{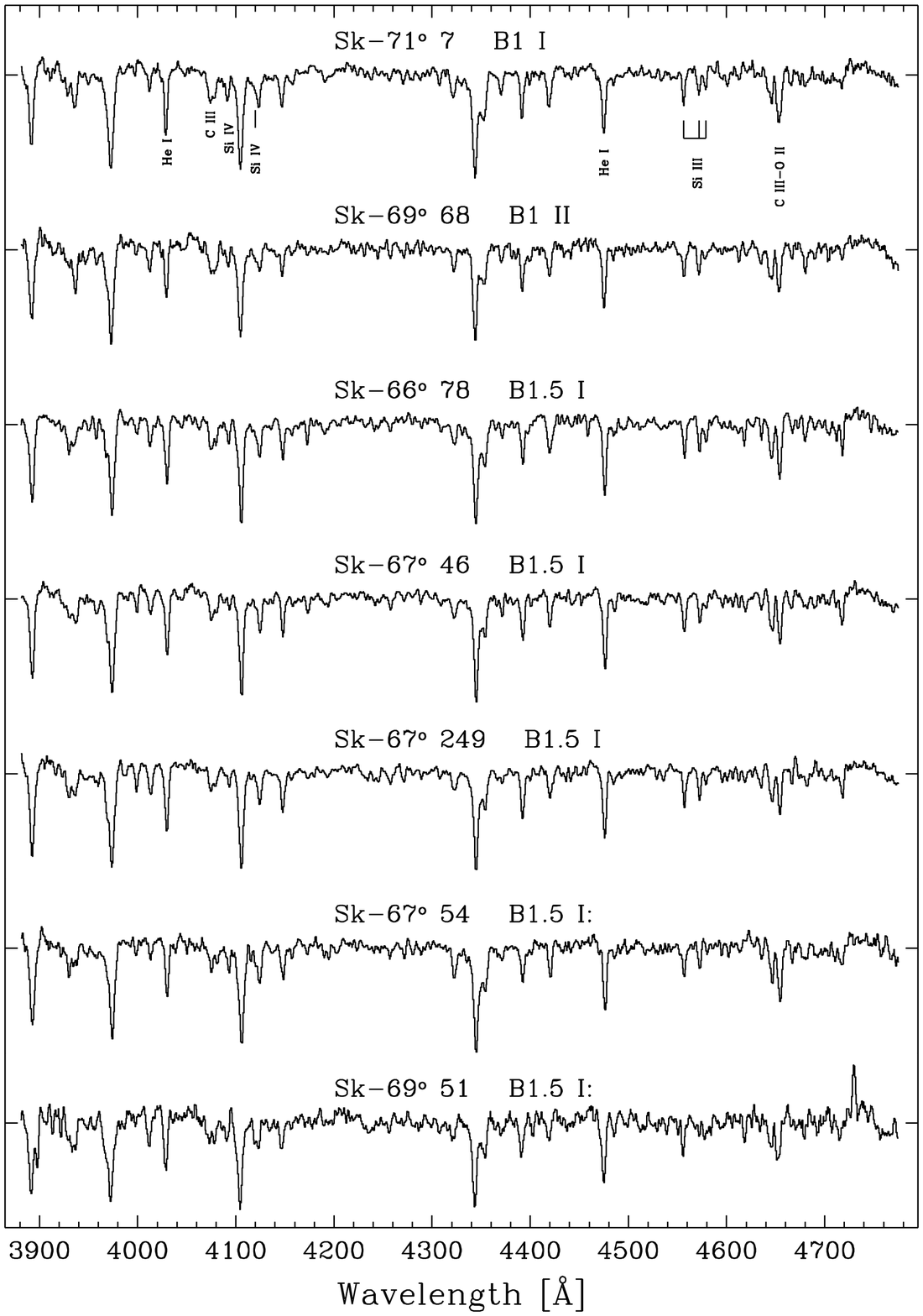}}
\end{figure}	     
		     
\begin{figure}	     
\centerline{\plotone{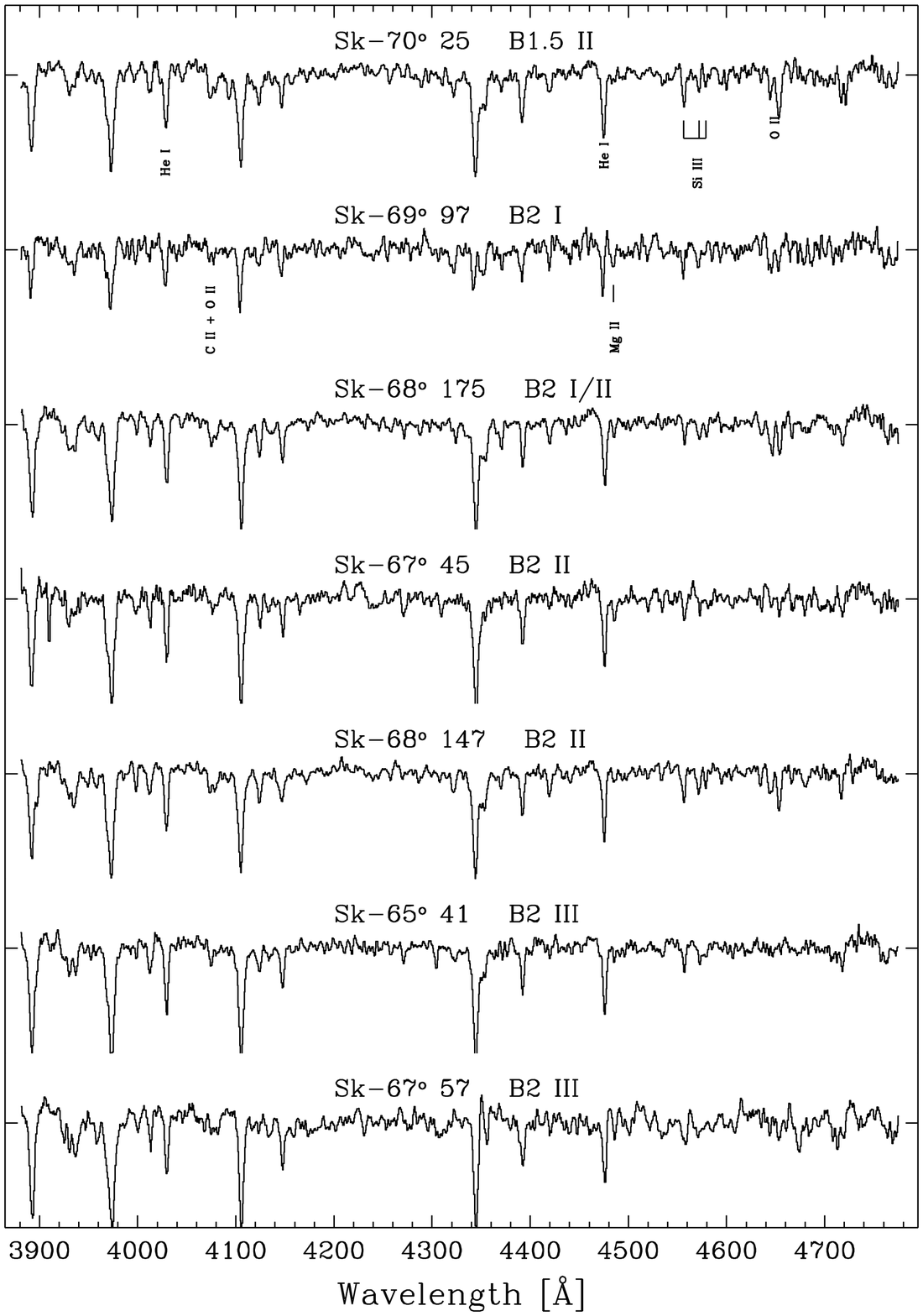}}
\end{figure}	     
		     
\begin{figure}	     
\centerline{\plotone{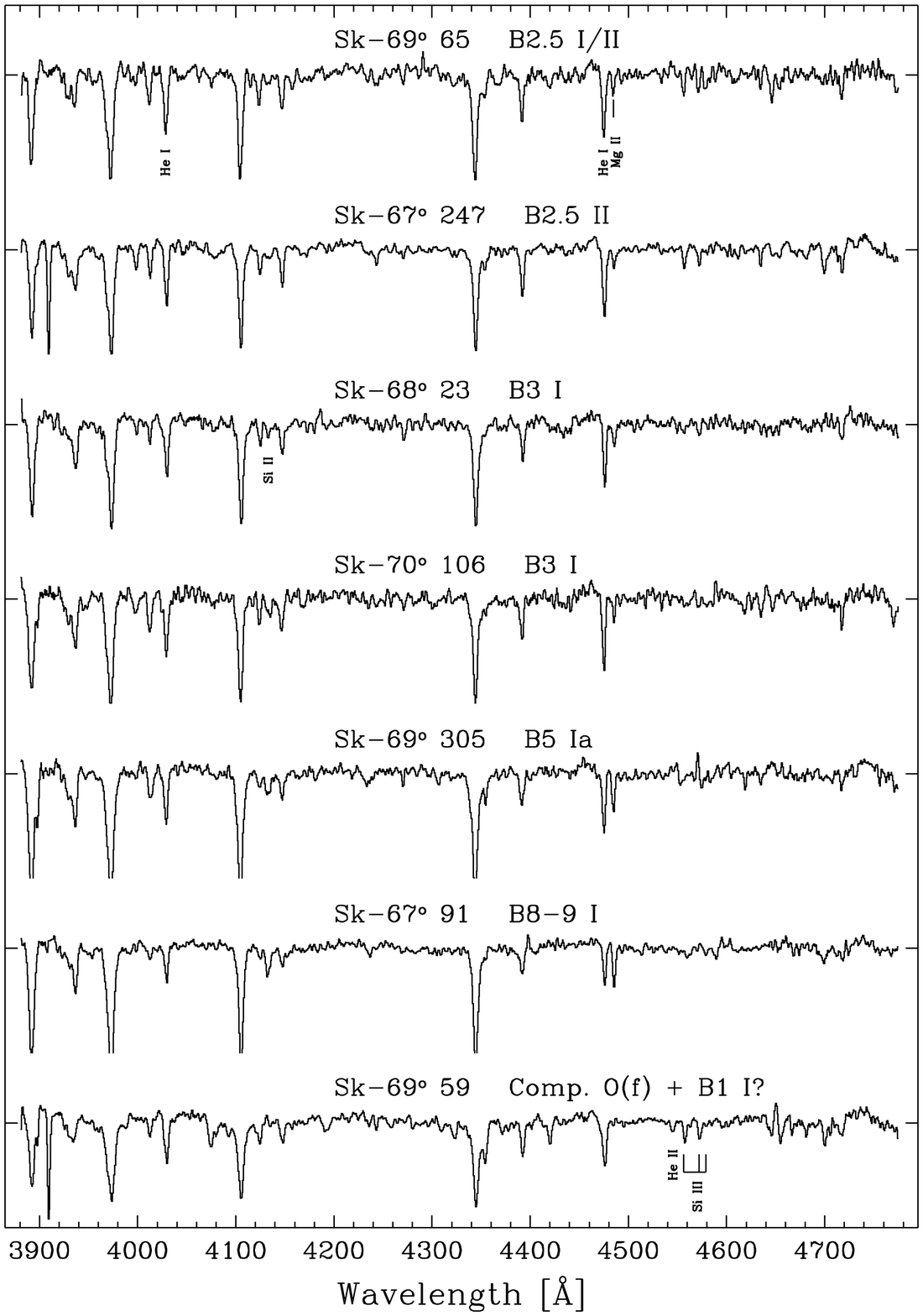}}
\end{figure}

\end{document}